\documentclass[conference,a4paper,10pt]{IEEEtran}

\usepackage[latin1]{inputenc}
\usepackage{times,amsmath}
\usepackage{amsfonts}
\usepackage{pstool}
\usepackage{multirow}
\usepackage{enumerate}
\usepackage{graphicx}
\usepackage{MnSymbol}
\usepackage{stfloats} 
\usepackage[table]{xcolor}
\usepackage[square, comma, sort&compress, numbers]{natbib}
\usepackage{nohyperref}
\usepackage{algorithm,algorithmic}
\usepackage{relsize}

\usepackage{float}
\usepackage[caption = false]{subfig}

\graphicspath{ {Figures/} }

\begin{document}

\title{RRH clustering and transmit precoding for interference-limited 5G CRAN downlink}

\author{
\IEEEauthorblockN{Muhammad Mahboob Ur Rahman\IEEEauthorrefmark{1}, Hadi Ghauch\IEEEauthorrefmark{1}, Sahar Imtiaz\IEEEauthorrefmark{1} and James Gross\IEEEauthorrefmark{1}}
\IEEEauthorblockA{\IEEEauthorrefmark{1}School of Electrical Engineering, KTH Royal Institute of Technology, Stockholm, Sweden \\\{mmurah, ghauch, sahari\}@kth.se, james.gross@ee.kth.se  }
}

\maketitle

\begin{abstract} 

In this work, we consider cloud RAN architecture and focus on the downlink of an antenna domain (AD) exposed to external interference from neighboring ADs. With system sum-rate as performance metric, and assuming that perfect channel state information is available at the aggregation node (AN), we implement i) a greedy user association algorithm, and ii) a greedy remote radio-head (RRH) clustering algorithm at the AN. We then vary the size of individual RRH clusters, and evaluate and compare the sum-rate gains due to two distinct transmit precoding schemes namely i) zero forcing beamforming (ZFBF), ii) coordinated beamforming (CB), when exposed to external interference of same kind. From system-level simulation results, we learn that in an interference-limited regime: i) RRH clustering helps, i.e., {\it cost-adjusted} performance when RRHs cooperate is superior to the performance when they don't, ii) for transmit precoding, the CB scheme is to be preferred over the ZFBF scheme. Finally, we discuss in detail the cost of RRH clustering, i.e., the piloting overhead (and the elements driving it), incorporate its impact on system sum-rate, and discuss its implications on the baseband processing capabilities of the RRHs. 

\end{abstract}


\section{Introduction}
\label{sec:intro}

Recently, with the commercialization of LTE standards, the discussions already began about what 5G will be like, both in terms of specifications and their enablers \cite{Andrews:JSAC:2014}. Though development efforts for 5G are in their nascent stage, it is becoming more evident that reducing the cell size, increasing the network density, and carefully managing the interference is the most natural and viable way-forward to increase the system data rates. 

Cloud Radio Access Network (CRAN) which is built upon the aforementioned design guidelines is a candidate architecture for 5G; there, the concept of cell is resolved altogether. Specifically, the serving transmit antennas, instead of being mounted on a single Base Station (BS) per cell, are grouped into multiple so-called Remote Radio Heads (RRH) which are then spread all over within an Antenna Domain (AD). These RRHs are then connected to a powerful computing node, so-called Aggregation Node (AN), via a fast backhaul.

In this work, we consider CRAN architecture and focus on the downlink of an AD exposed to external interference from neighboring ADs. The propagation scenario considered is of dense outdoor with Rician, line-of-sight (correlated) channels. With system sum-rate as performance metric, and assuming that perfect Channel State Information (CSI) is available at the AN, we implement i) a greedy user association algorithm, and ii) a greedy RRH clustering algorithm at the AN. We then vary the size of individual RRH clusters, and evaluate and compare the sum-rate gains due to two distinct transmit precoding schemes namely i) Zero-Forcing Beamforming (ZFBF), ii) (WMMSE based) Coordinated Beamforming (CB), when exposed to external interference of the same kind. From system-level simulation results, we learn that in an interference-limited regime, i) RRH clustering helps, i.e., {\it cost-adjusted} performance when RRHs cooperate is superior to the performance when they don't, ii) for transmit precoding, the CB scheme is to be preferred over the ZFBF scheme. We also discuss in detail the phenomena driving the cost of RRH clustering (i.e., the piloting overhead) which is either channel state information overhead, or, carrier synchronization overhead. We then use the piloting overhead as price to discount the system sum-rate and discuss its implications on the assumed baseband processing capabilities of the RRHs. 

The rest of this paper is organized as follows. Section-II summarizes the selected related work. Section-III discusses the system model and introduces the problem statement. Section-IV describes the approach taken in this paper. Section-V presents two greedy algorithms for user association and RRH clustering, implemented in this work. Section-VI presents some discussion about piloting overhead. Section-VII presents the performance evaluation results. Finally, section-VIII concludes the paper.

\section{Related work}
\label{sec:intro}

Going from single cell to multicell settings, interference has been widely recognized as the limiting factor on the system sum-rate performance. Coordinated Multipoint (CoMP) introduced the idea of cooperation among the BSs to mitigate intercell interference (though similar ideas under the so-called Network Multi-Input-Multi-Output (MIMO) were earlier investigated in~\cite{Karakayali:ICC:06}). In Joint Transmission (JT)  - one of the fundamental aspects of CoMP, multiple BSs coordinate their signals such that they align constructively, at each of the users, thereby turning interference from foe to friend~\cite{Gesbert:JSAC:10}. 

Naturally, all the aforementioned methods require distribution of perfect and global CSI (and occasionally, perfect RF carrier synchronization as well as users' data sharing) among all the base stations, which translates to great capacity and computational requirements for the backhaul. As a result, applying such coordination techniques to an entire network (i.e. turning all the BSs into a large virtual MIMO array) is infeasible. This in turn suggests that coordination has to be done locally, i.e., by forming clusters of neighboring BSs. 

Recognizing their potential and feasibility for real deployments, many works investigated the issue of BS clustering. The authors in~\cite{Papadogiannis:ICC:08} proposed a greedy dynamic clustering algorithm where ZFBF was considered, and the metric to maximize is the sum-rate  of the uplink. To reduce coordination overhead, a BS clustering algorithm was proposed in~\cite{Hong:JSAC:13} so that each user is served by only a small number of (potentially overlapping) BSs, i.e. by  jointly designing the BS clustering and the linear beamformers for all BSs in the network. Clustering and scheduling are jointly considered in~\cite{Gong:GC:11} to maximize the sum-rate and reduce the system complexity and overhead. There, the clusters are formed dynamically to reduce inter-cluster interference, and are allowed to be overlapped. Finally, the problem of joint optimization of the beamformer and the BS assignment (i.e. BS clustering) was considered in~\cite{Zhao:TWC:13}, where the aim is to minimize the per user requirement on the backhaul, subject to Quality-of-Service and per-BS power constraints. In a slightly different context, \cite{WeiYu:CWIT:2013} proposes various algorithms for static clustering of small cells in a two-tier network. 

To sum up the previous work, although themes such as coordination and clustering of transmit antennas have been extensively studied in the recent past, most of the earlier work focused on small scenarios (i.e., a small cluster of cooperating base stations), with very few works focusing on large deployments such as the one considered here. Moreover, virtually all previous coordination algorithms have been developed within the context of traditional cellular systems, where distributedness is a crucial aspect \cite{Bjornson:TSP:10}. However, the CRAN architecture (where a central compute node is available) considered in this work lifts the requirement of distributedness for a given algorithm, thereby changing the design paradigm. In other words, the availability of the global CSI at the central compute node renders the RRH clustering feasible for CRAN architecture which was otherwise considered as a daunting challenge for the traditional cellular systems. 

Nevertheless, RRH clustering in CRAN comes with yet another set of unique challenges, e.g., constrained backhaul, optimal splitting of baseband functionalities between the AN and the RRHs etc. \cite{Poor:WirelessComm:2015}. Having said this, the works on RRH clustering in CRAN downlink have started to emerge recently (see \cite{WeiYu:Access:2014} and the references therein). Specifically, \cite{WeiYu:Access:2014} considers the problem of joint user scheduling, user-centric (dynamic and static) clustering of RRHs and beamforming for the CRAN downlink where per-RRH backhaul constraints are explicitly considered in the network utility maximization problem. By approximating the non-convex, per-RRH backhaul constraints using reweighted $l_1$-norm technique, authors were able to solve the resulting weighted sum rate maximization problem through a generalized WMMSE approach.

However, this work, in addition to looking at the gain-cost trade-off for different levels of RRH clustering (which directly correspond to different available capacities at the backhaul): i) does the performance comparison of different transmit precoding schemes employed by the RRHs, ii) discusses in detail the baseband processing capabilities required by the RRHs, in the light of piloting overhead requirement of RRH clustering schemes.



\section{System Model and Problem Statement}
\label{sec:sys-model}

\subsection { System Model }

\begin{figure}[ht]
\begin{center}
	\includegraphics[width=3in,height=2.2in]{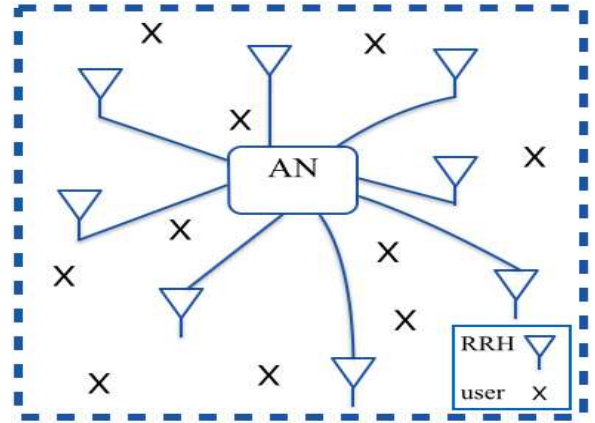} 
\caption{One Antenna Domain within the CRAN.}
\label{fig:das}
\end{center}
\end{figure}

We consider the CRAN architecture where $N$ $M$-antenna Radio-Frequency (RF) transceivers, the RRHs, are widely spread out within a square AD (see Fig. \ref{fig:das}). Furthermore, all the $N$ RRHs are connected to the central controller, the AN, via a delay-free, error-free, high-throughput (yet constrained) wired/wireless backhaul. Most of the precoding/baseband processing takes place at the AN while the RRHs mainly act as RF front ends. We then focus on downlink transmissions, i.e., the AN needs to serve $K$ single-antenna users present in the AD via available resources, i.e., $N\times M$ transmit antennas, during every downlink time-slot. 

We next summarize all the important assumptions this work will build upon. Perfect and global instantaneous CSI is available at the AN. The system is time-slotted with $T$ sec long time-slots. The AN is in a saturating traffic condition, i.e., there is always data sitting in buffers at the AN which needs to be scheduled to the users. We assume a dense outdoor deployment, i.e., $N$ and $K$ are both very large; additionally, each AD receives external interference from its neighboring ADs (upto 8, according to the square grid); therefore, the scenario under consideration is severely interference-limited. Furthermore, for any two users in close proximity of each other, their Rician flat fading channels (w.r.t. a common transmit antenna) are correlated. Finally, all the users in the AD are stationary.

\subsection { Problem Statement }

In this work, we use the sum-rate as the performance metric of the system:

\begin{equation}
	\label{eq:opt_sumrate}
	R_{\sum}=\sum_{k=1}^K \log_2 ( 1+\gamma_{k} )
\end{equation}

where $\gamma_{k}$ is the Signal to Interference-plus-Noise Ratio (SINR) at user $k$, and is defined as (assuming unit-energy information symbols):
\begin{equation}
	\label{eq:opt_sumrate}
	\gamma_{k} = \frac{P_k \bigg| [ \mathbf{H} \mathbf{W} ]_{k,k} \bigg|^2 } { \sum\limits_{j=1,j \neq k}^{K} P_j \bigg| [ \mathbf{H} \mathbf{W} ]_{k,j} \bigg|^2 + \sum\limits_{i=1}^{K_{out}} P_i \bigg| [ \mathbf{H_{cross}} \mathbf{W_{out}} ]_{k,i} \bigg|^2 + \sigma^2 }
\end{equation}

Here, $P_x$ is the power allocated to user $x$; $\mathbf{H}$ $\in$ $\mathbb{C}^{K\times (M \times N)}$ is the channel between the RRHs and the users which are inside the AD, $\mathbf{W}$ $\in$ $\mathbb{C}^{(M \times N)\times K}$ is the precoding matrix used by the RRHs inside the AD; $\mathbf{H_{cross}}$ $\in$ $\mathbb{C}^{K\times (M_{out} \times N_{out})}$ is the cross channel, between the RRHs outside the AD and the users inside the AD; $\mathbf{W_{out}}$ $\in$ $\mathbb{C}^{(M_{out} \times N_{out})\times K_{out}}$ is the precoding matrix used by the RRHs outside the AD; $\sigma^2$ is the noise power. Finally, note that the first term in the denominator represents internal (intra-AD) interference, while the second term in the denominator represents the external (inter-AD) interference.

Then, to compute the sum-rate, we employ a sequential, heuristic, sub-optimal approach. Specifically, we implement a greedy user association algorithm, followed by a greedy RRH clustering algorithm, followed by a transmit precoding scheme, to compute the system sum-rate.
\section { Proposed Approach: Let RRHs Cooperate }
\label{sec:ourapp}

At the beginning of every downlink transmission slot, the AN needs to solve a number of sub-problems, i.e., user selection, user association, transmit antenna (RRH) clustering, transmit precoding, transmit power control, interference coordination etc.

In this work, we assume that $K \le M \times N$ and $\mathrm{rank}(\mathbf{H})=K$ (hence, there is no user selection). For user association and transmit antenna (RRH) clustering, we will present two CSI-based greedy algorithms in section-V. For transmit precoding, we employ two low-complexity linear schemes: i) ZFBF, ii) WMMSE based CB \cite{ZQLuo:TSP:2011}. Finally, transmit power loading could be achieved by standard water-filling algorithms \cite{Palomar:TSP:2005} (not implemented in this work though). 

We are then interested to evaluate the gain in system performance due to inter-RRH cooperation (a.k.a RRH clustering), and and then trade it against the cost of cooperation. In this paper, two or more RRHs are said to cooperate if they are able to do {\it joint} ZFBF based (CB based) transmit precoding to jointly (individually) serve their associated users. Thus, inter-RRH cooperation necessarily requires certain piloting overhead (driven by either carrier synchronization overhead, or, CSI acquisition overhead). Therefore, from now onwards, the cooperation cost refers to piloting overhead which will be discussed in more detail in section-VI.

Next, we identify three distinct cases of inter-RRH cooperation and intend to evaluate the system sum-rate performance for each case. Broadly speaking, the expectation is that more cooperation implies more performance and more (piloting) overhead\footnote{In case of constrained backhaul, the cluster size $B$ of an individual cluster is dictated by the capacity of the backhaul (not considered in this work though).}, and vice versa. We now discuss all the three cases one by one.

\subsection{Global Coordination (GC): All RRHs Cooperate}

For the ZFBF based transmit precoding, all the RRHs form one large virtual antenna array via RF carrier synchronization mechanism. Moreover, since all the $N$ RRHs jointly serve all the $K$ users, the user association algorithm need not be run. The AN then inverts the global channel matrix $\mathbf{H}$ to generate the global precoding matrix $\mathbf{W}$. With this, the interference is totally eliminated within the AD. 

For the CB based transmit precoding, first $J \le M$ users are associated with every RRH (via the user association algorithm described in section-V). Then, each RRH computes its transmit precoder by means of the algorithm proposed in \cite{ZQLuo:TSP:2011}.

\subsection{Local Coordination (LC): Neighboring RRHs Cooperate}

In this case, first $J \le M$ users are associated with every RRH (assuming users are uniformly distributed within the AD). Then, the AD is partitioned into $C$ disjoint clusters with $B$ RRHs each. Then for ZFBF based precoding, within each RRH cluster, the AN inverts the local channel matrix to generate a local precoding matrix. On the other hand, for the CB based precoding, each RRH computes its transmit precoder by means of the algorithm proposed in \cite{ZQLuo:TSP:2011}. The main motivation for the LC scheme, compared to GC scheme, is to reduce the piloting overhead. This overhead reduction, however, comes at a cost of inter-cluster interference.

\subsection{No Coordination (NC): RRHs don't Cooperate}

This case is applicable to the ZFBF scheme only. In this case, first $J \le M$ users are associated with every RRH; each RRH then serves its associated users locally, i.e., RRHs don't cooperate at all; rather, each RRH computes its local precoding matrix via the ZFBF scheme. This case results in lowest overhead and lowest performance (as there is inter-RRH interference now present, which is much stronger than the inter-cluster interference in the LC scheme).

\section { The Implemented Algorithms }
\label{sec:algo}

\subsection{User Association Algorithm}
A user association (UA) algorithm assigns $J \le M$ orthogonal users to each RRH. In this paper, we have implemented a simple, CSI-based, RRH-centric, UA algorithm which is summarized below:

\begin{enumerate}

\item Select an RRH which is not considered so far.

\item Pick the $J$ users from the '{\it unassigned users}' set which have most favorable channel conditions w.r.t. the considered RRH.

\item Repeat (1) and (2) until unassigned user set $=\{\}$.

\end{enumerate}

\subsection{RRH Clustering Algorithm}
An RRH clustering algorithm, as expected, partitions the AD into multiple disjoint clusters. In this work, we have implemented a simple, CSI-based, greedy RRH clustering method (inspired by the work in \cite{Papadogiannis:ICC:08}) which is summarized below:

\begin{enumerate}

\item Choose number of clusters $C$ (each of size $B$) to create.

\item Start forming a new RRH cluster: 

\begin{enumerate}

\item Pick an RRH along with its associated users from the '{\it unassigned RRH-plus-users}' set, which, when combined with RRHs in the '{\it RRH-plus-users selected for current cluster}' set, results in maximum sum rate. 

\item Repeat (2a) until cluster size reaches $B$. 

\end{enumerate}

\item Repeat (2) until all $C$ clusters are formed and no RRH is left unclustered.

\end{enumerate}

\section { The Piloting Overhead }
\label{sec:overhead}

As is briefly mentioned before, the cooperation among RRHs comes at a cost: the piloting overhead which refers to the time resources (symbols) consumed on the downlink for training. Piloting overhead could be driven by either the CSI acquisition overhead, or, carrier synchronization overhead. The CSI acquisition overhead itself is then driven by the users'/scatterers' speed, while the carrier synchronization overhead is driven by the stability of oscillators employed by RRHs and users. 

At this stage, it is worth mentioning that the ZFBF based RRH clustering schemes (GC and LC) require both carrier synchronization as well as CSI at the AN, while the CB based RRH clustering schemes (GC and LC) require only the CSI at the AN. However, since we assume stationary users in this work, for the ZFBF scheme, the piloting overhead is then mostly driven by the carrier synchronization overhead. On the other hand, due to stationary users assumption, for the CB scheme, the piloting overhead, if present, comes due to movement of the scatterers within the AD. Having this said, the mechanisms for CSI acquisition are well-known in the literature. Therefore, we will only discuss the specifics of the carrier synchronization overhead under CRAN architecture in the rest of this section.

Precisely speaking, for every RRH cluster and during every downlink timeslot, near-perfect frequency, phase and timing synchronization among RRHs are needed, in order for them to do a joint ZFBF transmission successfully to their associated users. 

Timing synchronization is the least stringent requirement of all, and can be fulfilled with relative ease; thanks to the cyclic prefix property of orthogonal frequency division multiplex (OFDM) symbols, and thanks to Line-of-Sight (LOS) channels which greatly reduce the delay spread of the signal.

Frequency synchronization requires the AN to periodically broadcast a pilot/training/reference signal to the RRHs on the backhaul link. Both the duration $T_{est}$ of training interval and the rate $\frac{1}{T_{slot}}$ (training intervals per sec) mainly depend upon the Signal to Noise Ratio (SNR) $\gamma$ of the backhaul link, i.e., $\frac{T_{slot}}{T_{est}} \approx k\sqrt{\gamma}$ \cite{Quitin:TWC:2013}. 

Since in the considered CRAN architecture, (wireless) backhaul link and downlink channel operate on different frequencies, RRHs operate in Frequency Division Duplex (FDD) mode. Then, for both uplink and downlink data forwarding, RRHs could assume either Amplify-and-Forward (AF) operation, or, Decode-and-Forward (DF) operation. In either case (AF/FDD or DF/FDD), explicit frequency offset estimation plus frequency correction operations are required at each RRH, for cooperative downlink transmissions \cite{Mahboob:VTCS:2015}. In this work we assume that the RRHs operate in DF/FDD mode, i.e., {\it we assume that RRHs have limited baseband processing capability available}.

Phase synchronization is perhaps the most challenging of all. Specifically, let $N_{c_i}$ represent the number of RRHs and $K_{c_i}$ represent the number of users in cluster $c_i$. Then, at time $t=0$, one needs to estimate $N_{c_i}\times K_{c_i}$ pair-wise phase offsets between each RRH and each user, for cluster $c_i$. Therefore, we need to estimate a total of $\sum_{i=1}^C N_{c_i}\times K_{c_i}$ unknown parameters ($C$ is the total number of clusters). Then, assuming that estimation of each (pair-wise) unknown requires only one training symbol, the total {\it start-up/initialization cost} is: $\sum_{i=1}^C K_{c_i}=K$ training symbols. 

Next, the pairwise phase offsets randomly drift over time due to Brownian motion phase drift \cite{Quitin:TWC:2013}. Therefore, periodic re-estimation of all $\sum_{i=1}^C N_{c_i}\times K_{c_i}$ phase offsets is necessary. It is well known that, similar to fading channels, there is a coherence interval for phase offsets as well which depends on oscillator stability as well as the objective function (e.g., ZFBF condition). Let $PF$ represent the piloting frequency, then $\frac{1}{PF}$ represents the phase offset coherence interval, then $PF \times (\sum_{i=1}^C N_{c_i}\times K_{c_i})$ will be the number of phase offset values which need to be estimated per second. This gives rise to the {\it cluster maintenance cost}. 

Again, it is worth mentioning that pairwise phase offsets are easiest to estimate by RRHs themselves when users send training/pilot sequences during the uplink transmission phase. Hence again, we assume that RRHs have limited baseband processing capability available.

To recap, the total phase synchronization cost is: $\Omega=K \times PF$ (training symbols per second). Then, our {\it cost-adjusted} system sum-rate (for both local and global coordination of RRHs) is \cite{RHeath:TWC:2012}:
\begin{equation}
	\label{eq:adj_sumrate}
	R_{\sum}=\bigg(\frac{W-\Omega}{W}\bigg)\sum_{k=1}^K \log_2 (1+\gamma_{k}(\Omega))
\end{equation}
where $W$ is the total number of symbols per second, for the downlink. One also needs to be aware of the fact that SINR $\gamma_k(\Omega)$ itself has a direct relation with the total synchronization overhead $\Omega$ (not investigated in this work though).

\section{Performance Evaluation}
\label{sec:results}

\subsection{Simulation Setup}

During every experiment run, we drop $N=24$ RRHs (with $M=4$ antennas each), and $K=48$ users according to uniform distribution over a square AD of length $0.25$ km (which corresponds to an inter-site distance of 50 meters). 

To realize a realistic dense outdoor scenario, we modeled all the channels in the AD as spatially correlated Rician fading channels. To implement fading and path-loss, we used the following model \cite{Papadogiannis:ICC:08}:
\begin{equation}
	\label{eq:channel_model}
	h_{mn;k} = \frac{ \eta_{mn;k} \sqrt{\zeta_{mn;k}} } { \sqrt{d_{mn;k}^{\alpha}} }  
\end{equation}
where $h_{mn;k}$ is the channel gain between $m$-th antenna of RRH $n$ and user $k$, $\eta_{mn;k}$ represents small-scale fading, $\zeta_{mn;k}$ represents shadow fading, $d_{mn;k}^{\alpha}$ represents path-loss, $d_{mn;k}$ is the distance, ${\alpha}$ is the path-loss exponent. Specifically, $\eta_{mn;k}=\Re_{mn;k}+j\Im_{mn;k}$ where  $\Re_{mn;k}\sim \mathcal{N}(\mu,\sigma)$ and $\Im_{mn;k}\sim \mathcal{N}(0,\sigma)$; $\mu=\sqrt{\mathcal{K}/(\mathcal{K}+1)}$, $\sigma=\sqrt{1/(2\mathcal{K}+2)}$ where $\mathcal{K}$ is the Rician factor which was set to 1. Moreover, $\zeta_{mn;k}\sim \mathcal{N}(0dB,8dB)$; ${\alpha}=3.76$. Furthermore, for the path-loss, we employed the 3GPP-LTE path-loss model \cite{Emil:PIMRC:2011}:
\begin{equation}
	\label{eq:pathloss_model}
	PL_{mn;k} (dB)= 36.3 + 37.6\log_{10} d_{mn;k} (m)
\end{equation}

Next, to implement spatial correlation between the channels, classical Kronecker product model was adopted \cite{Kermoal:JSAC:2002}. Remember that in Kronecker product model, the Tx-side correlation matrix $\mathbf{R}_{Tx}$ (for correlation between transmit antennas) is decoupled from the Rx-side correlation matrix $\mathbf{R}_{Rx}$ (for correlation between receive antennas). Then, to generate the $N$ block-diagonal entries $\mathbf{R}_{Tx,n} \in \mathbb{R}^{M\times M}$ of $\mathbf{R}_{Tx} \in \mathbb{R}^{(N\times M)\times (N\times M)}$, we employed the exponential correlation model \cite{Loyka:CL:2001}:
\begin{equation}
	\label{eq:R_Tx_diag}
	[\mathbf{R}_{Tx,n}]_{p,q} = \rho_t^{ |p-q| } 
\end{equation} 
where $0\le \rho_t \le 1$ is the parameter which controls the amount of correlation, $p$ and $q$ are antenna indices. Moreover, since in our case, RRHs are widely spread out within AD, to fill-in the off-block-diagonal entries of $\mathbf{R}_{Tx}$,we used the following exponential-like model:
\begin{equation}
	\label{eq:R_Tx_off_diag}
	[\mathbf{R}_{Tx}]_{i,j} = \rho_t^{' \lceil d_{ij}/d_{min} \rceil } 
\end{equation} 
where $0\le \rho_t^{'} \ll \rho_t$ is the parameter which controls the amount of correlation; $d_{min}$ is the minimum pair-wise inter-RRH distance, for a given deployment of RRHs; $d_{ij}$ is the distance between RRH $i$ and RRH $j$. 

Similarly, we generated $\mathbf{R}_{Rx} \in \mathbb{R}^{K\times K}$ as follows:
\begin{equation}
	\label{eq:R_Rx}
	[\mathbf{R}_{Rx}]_{i,j} = \rho_r^{\lceil d_{ij}/d_{min} \rceil } 
\end{equation}
where $0\le \rho_r \le 1$ is the parameter which controls the amount of correlation; $d_{min}$ is the minimum pair-wise inter-user distance, for a given deployment of users; $d_{ij}$ is the distance between user $i$ and user $j$. 
Throughout the simulations, $\rho_t=\rho_r=0.5$ and $\rho_t^{'}=\rho_t^M$ was used.

\begin{figure}[ht]
\begin{center}
	\includegraphics[width=2in,height=1in]{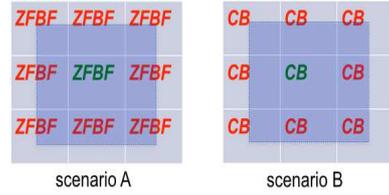} 
\caption{ out-of-AD (external) interference modeling }
\label{fig:outofAD_int}
\end{center}
\end{figure}

Furthermore, in reality, there is additional out-of-AD interference which affects the transmission performance of RRHs within the AD. Therefore, as our initial attempt to model out-of-AD (external) interference, we deployed 8 neighboring ADs around the periphery of the AD under investigation (see Fig. \ref{fig:outofAD_int}). We then dropped $N_{out}=3N$ additional RRHs with $M_{out}=M$ antennas each, and $K_{out}=(N_{out}\times M_{out})/2$ additional users in the blue shaded region of the 8 neighboring ADs (see Fig. \ref{fig:outofAD_int}). We then identify two distinct simulation scenarios based upon transmit precoding scheme used: i) scenario A: inside-AD RRHs and out-of-AD RRHs both employ ZFBF scheme, ii) scenario B: inside-AD RRHs and out-of-AD RRHs both employ WMMSE based CB scheme. Finally, we set power allocation per inside-AD RRH to the same value as the power allocation per out-of-AD RRH (i.e., no water-filling was done).

\subsection{Simulation Results}

Broadly speaking, we first switched external interference on and off to compare scenario A against scenario B. We then decided to go with scenario B and keep the external interference always on. 

\subsubsection{External interference is switched on and off and scenario A is compared against scenario B} 

\begin{figure}
\subfloat[transmit precoding is ZFBF \label{fig:main_resulta}]
{\includegraphics[width = 3in,height=2.2in]{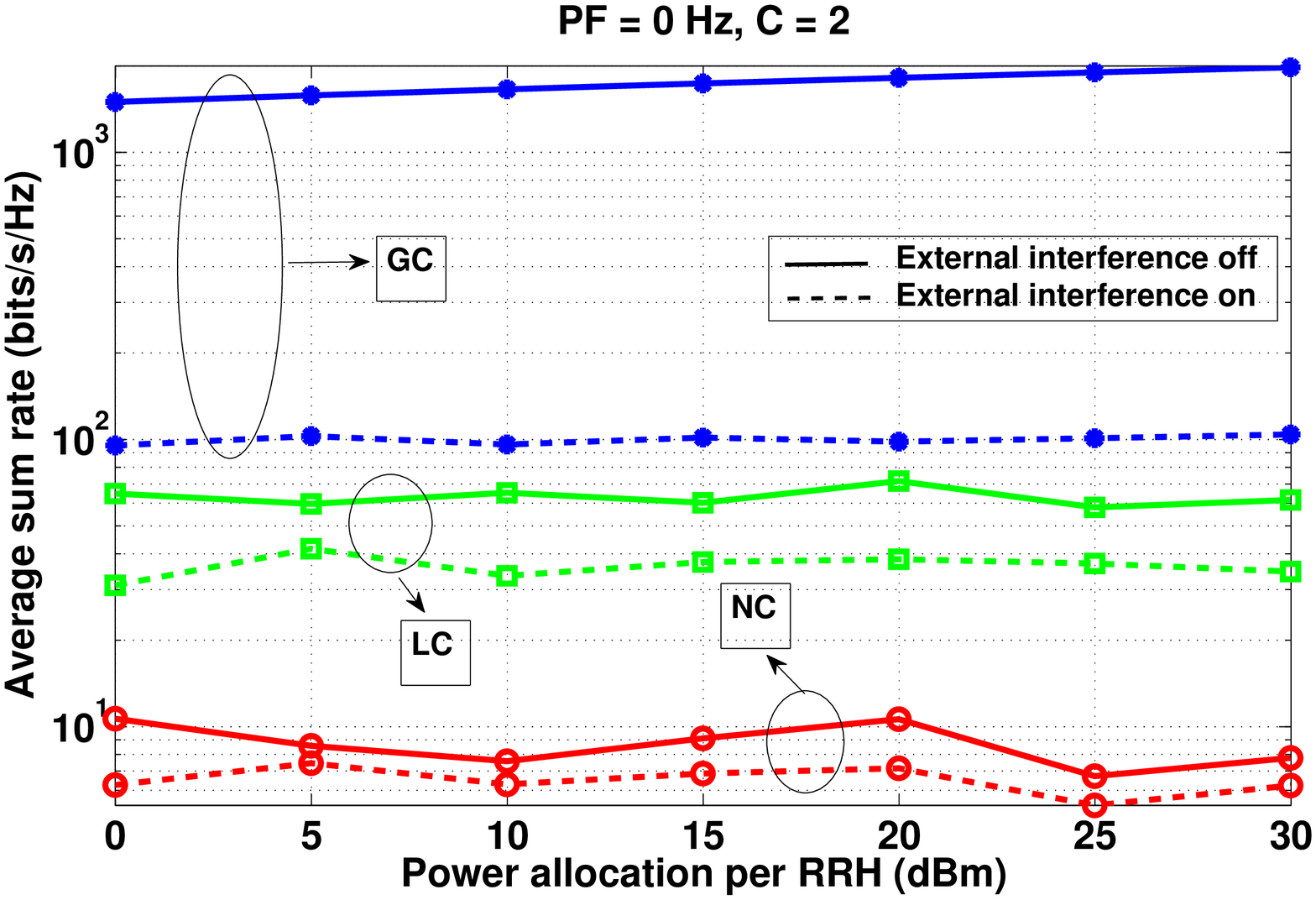}} \\
\subfloat[transmit precoding is WMMSE based CB \label{fig:main_resultb}]
{\includegraphics[width = 3in,height=2.2in]{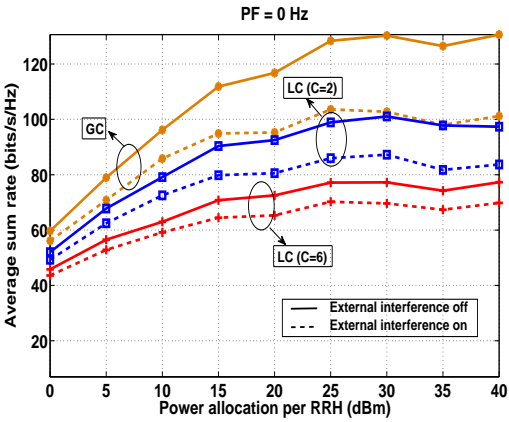}}\\
\caption{Ergodic sum-rate vs. power allocation per RRH }
\label{fig:main_result}
\end{figure}

Fig. \ref{fig:main_result} plots the average sum-rate from Equation (\ref{eq:adj_sumrate}) against power allocation per RRH where the external interference is first switched off and later is switched on. Specifically, Fig. \ref{fig:main_resulta} (Fig. \ref{fig:main_resultb}) represents the scenario A (scenario B). First and foremost, from Fig. \ref{fig:main_resulta}, we  learn that the RRH clustering scheme (both GC and LC schemes) performs better than the NC scheme (where each RRH serves its users independent of others). Moreover, the performance advantage of RRH clustering still remains there even if we take into account low-to-medium piloting overhead (either due to users' mobility, or, instability of users' oscillators).

Additionally, Fig. \ref{fig:main_result} leads us to several other observations: i) GC scheme in scenario A is most sensitive to external interference with an order of magnitude drop in sum-rate, ii) in both scenarios, GC scheme always outperforms LC scheme (though the performance gap is much narrower in scenario B), iii) in both scenarios, small cluster sizes (or, more number of clusters) decrease the system performance and vice versa, iv) scenario A is quite insensitive to power allocated to RRHs for the full range of SNR values, while the performance of scenario B can be tuned to any desired value by transmit power allocation (though it also saturates at high SNR values). 

Coming to explanations of above observations, both ii) and iii) are due to the fact that external interference has the same adverse affect on both GC and LC schemes; therefore, it is actually the inter-cluster interference which causes the LC scheme to perform worse than the GC scheme. For iv), one potential explanation is the following. For scenario A, due to ZFBF being aggressive in allocating power to RRHs, an increase in transmit power allocation will most likely result in proportional increase in both external interference as well as inter-cluster interference. Therefore, the net effect is no increase in system sum-rate with increase in transmit power allocation. This prompts us to suggest that for scenario A, lesser power should be allocated to RRHs. On the other hand, for scenario B, an increase in power allocation results in an increase in sum-rate; this is most likely due to the fact that the CB scheme controls transmit power (and hence manages interference) more efficiently. 

Now, recall that the ZFBF scheme assumes the following: $K \le M \times N$ and $\mathrm{rank}(\mathbf{H})=K$. The CB scheme, on the other hand, removes such restrictions on channel condition and maximum number of served users. This fact coupled with the observations i), ii) and iv) from Fig. \ref{fig:main_result} prompts us to go for scenario B for the rest of this section. Additionally, we turn on the external interference once and for all, for the rest of this section.

\subsubsection{External interference is always switched on and precoding scheme is WMMSE based CB} 

\begin{figure}[ht]
\begin{center}
	\includegraphics[width=3in,height=2.2in]{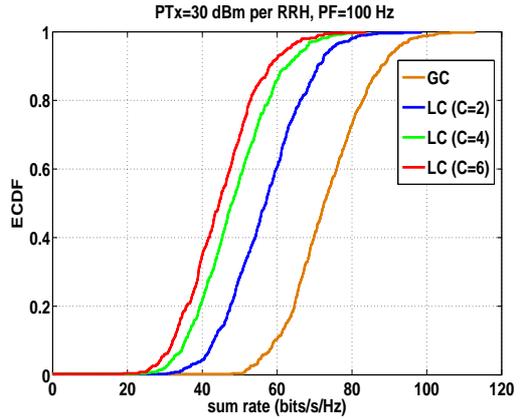} 
\caption{ ECDF of sum-rate }
\label{fig:ecdf}
\end{center}
\end{figure}

Fig. \ref{fig:ecdf} plots the Empirical Cumulative Distribution Function (ECDF) of the system sum-rate. The figure basically corroborates the observations ii) and iii) from Fig. \ref{fig:main_result}.

\begin{figure}[ht]
\begin{center}
	\includegraphics[width=3in,height=2.2in]{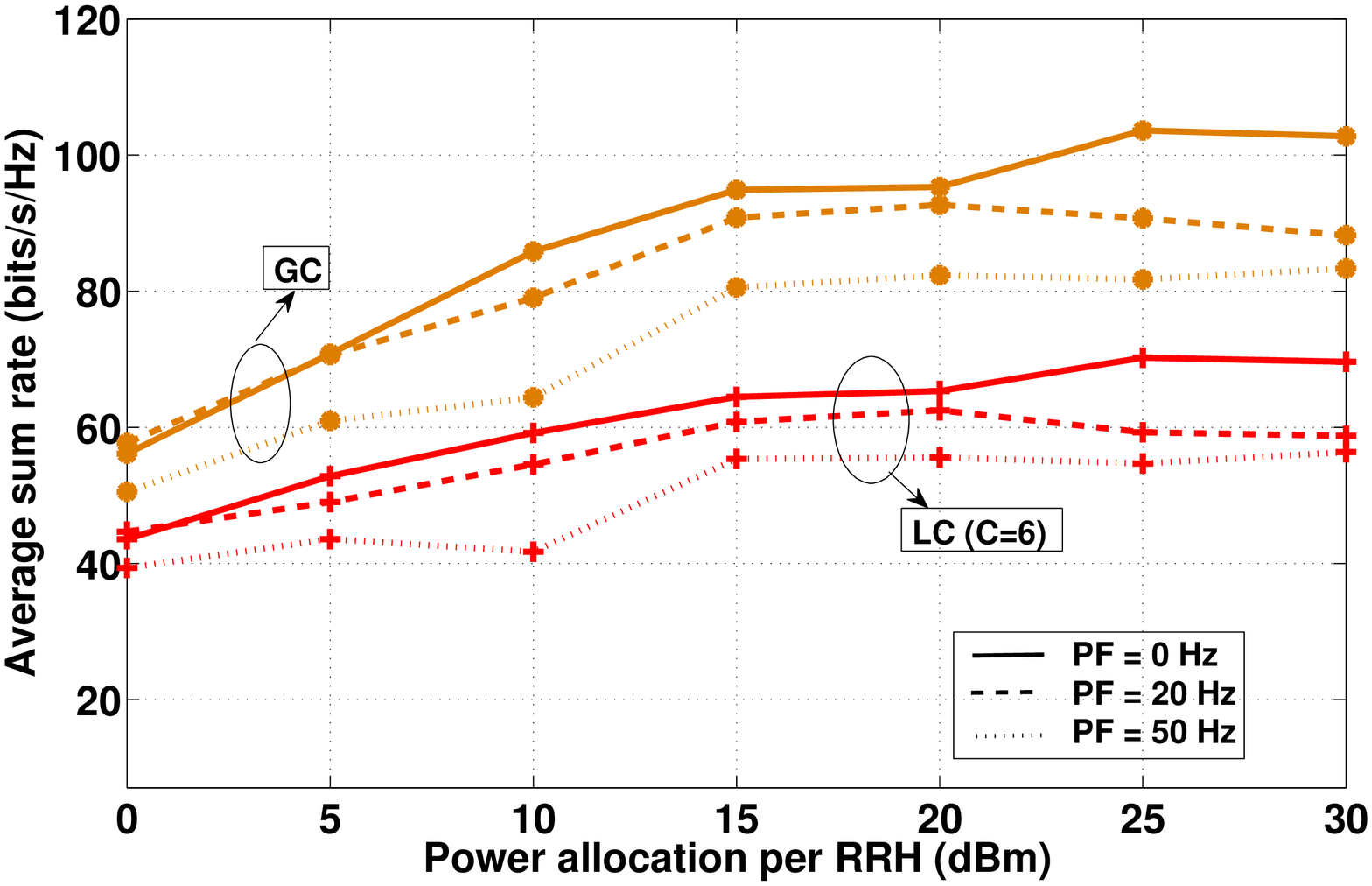} 
\caption{ Effect of piloting overhead on ergodic sum rate }
\label{fig:RvsP_Fcoh}
\end{center}
\end{figure}

Fig. \ref{fig:RvsP_Fcoh} plots the average sum-rate against per RRH transmit power allocation for different piloting frequencies. From the figure, it is evident that the piloting overhead greatly impacts (degrades) the sum-rate of both GC and LC schemes.

\begin{figure}[ht]
\begin{center}
	\includegraphics[width=3in,height=2.2in]{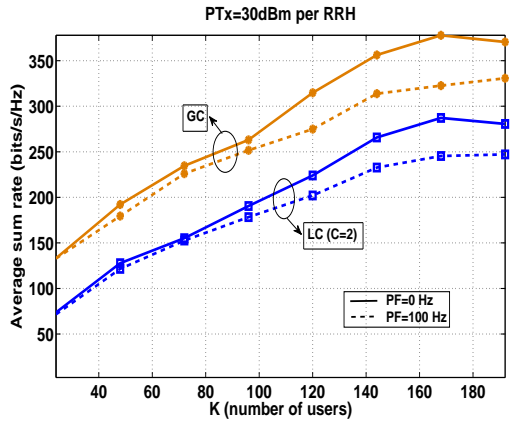} 
\caption{ Ergodic sum rate vs. K (number of users) }
\label{fig:avg_sum_rate_vs_K}
\end{center}
\end{figure}

Fig. \ref{fig:avg_sum_rate_vs_K} plots the average sum-rate behavior when number of served users within the AD is varied. For this plot, we used $N=24$ but set $M=8$. From the figure, we learn that an increase in number of users results in an increase in system sum-rate. This in turn signifies the efficiency of the CB as transmit precoding scheme which has turned a severely interference-limited deployment into (somewhat) interference-controlled situation. Therefore, the CB scheme offers the advantage of scalability of system sum-rate in terms of number of served users.

\section{Conclusion}
\label{sec:conclusion}

In this work, we investigated the (sum-rate) gain vs. (piloting) cost tradeoff of RRH clustering for the downlink of Cloud RAN in an interference-limited, dense outdoor setting. Specifically, we implemented a greedy RRH clustering algorithm, varied the size of individual RRH clusters, and compared the performance of two distinct transmit precoding schemes namely i) ZFBF, ii) WMMSE based CB, when exposed to external interference of the same kind. System-level simulation results suggested that in an interference-limited regime: i) RRH clustering helps, ii) for transmit precoding, the CB scheme is to be preferred over the ZFBF scheme. We also discussed in detail the cost of RRH clustering, i.e., the piloting overhead (and the elements driving it), incorporated its impact on system sum-rate, and discussed its implications on the baseband processing capabilities of the RRHs. Immediate future work will involve evaluating the gain vs. cost trade-off of RRH clustering for the case of mobile users.

\appendices

\section*{Acknowledgements}
KTH gratefully acknowledge the funding and support from Huawei, Finland.

\footnotesize{
\bibliographystyle{IEEEtran}
\bibliography{references}
}

\vfill\break

\end{document}